# A Piecewise Approach for the Analysis of Exact Algorithms

**Katie Clinch** ✉
School of Computer Science and Engineering, UNSW Sydney, Australia

**Serge Gaspers** ✉
School of Computer Science and Engineering, UNSW Sydney, Australia

**Zixu He** ✉
School of Computer Science and Engineering, UNSW Sydney, Australia

**Abdallah Saffidine** ✉
School of Computer Science and Engineering, UNSW Sydney, Australia

**Tiankuang Zhang** ✉
School of Computer Science and Engineering, UNSW Sydney, Australia

— **Abstract** —————————————————————————

To analyze the worst-case running time of branching algorithms, the majority of work in exponential time algorithms focuses on designing complicated branching rules over developing better analysis methods for simple algorithms. In the mid-2000s, Fomin et al. [9] introduced measure & conquer, an advanced general analysis method, sparking widespread adoption for obtaining tighter worst-case running time upper bounds for many fundamental NP-complete problems. Yet, much potential in this direction remains untapped, as most subsequent work applied it without further advancement.

Motivated by this, we present piecewise analysis, a new general method that analyzes the running time of branching algorithms. Our approach is to define a similarity ratio that divides instances into groups and then analyze the running time within each group separately. The similarity ratio is a scale between two parameters of an instance $I$. Instead of relying on a single measure and a single analysis for the whole instance space, our method allows us to take advantage of different intrinsic properties of instances with different similarity ratios.

To showcase its potential, we reanalyze two 17-year-old algorithms from Fomin et al.[6] that solve 4-Coloring and #3-Coloring respectively. The original analysis in their paper gave running times of $\mathcal{O}(1.7272^n)$ and $\mathcal{O}(1.6262^n)$ respectively for these algorithms. Our analysis improves these running times to $\mathcal{O}(1.7207^n)$ and $\mathcal{O}(1.6225^n)$.

## 1 Introduction

Branching algorithms have been widely used for more than 50 years to solve NP-complete problems. To achieve better worst-case running time upper bounds, there are two main approaches. One is to design smarter algorithms that exploit the input's properties and make a large case distinction. The other approach is to improve the analysis of a simple (existing) algorithm. The first approach had been widely used in the past, but its effectiveness is often hamstrung by the design and the analysis of the algorithm being too intertwined. This means that such algorithms typically do not benefit from advances in algorithm analysis techniques. Even for simple algorithms, we typically do not know what the true worst-case running time is. Often the known upper and lower bounds do not match. Therefore, a major issue in the field of exponential time algorithms is the development of better methods to analyze algorithms, in particular branching algorithms.

In the mid-2000s, Fomin et al. introduced measure & conquer (M&C) in [10], after some preliminary work had been done in the SAT community (see, e.g., [17]) and the quasiconvex analysis by Eppstein [5] in relation to graph coloring algorithms. Before this, in order to obtain a running time upper bound with respect to a variable $n$, one would track the progress of a branching algorithm in terms of $n$ and solve a set of recurrences resulting from this analysis. M&C introduced the use of a potential function, a so-called measure, to track the progress of the algorithm. A measure is a weighted linear function of parameters of the instance; parameters such as the numbers of vertices of various degrees or the size of a solution. One then needs to minimize an upper bound on the measure with respect to $n$, subject to convex constraints arising from the analysis of the branching rules [14]. In [10], Fomin et al. used a linear function where all coefficients and weights are positive. Later, several authors expanded M&C to a wider set of measures: negative coefficients were used by [8], a small [20] or large [16] number of potentials were used that add terms to the measure conditional on properties of the instance or the algorithm, logarithmic terms were used by [15] and [13] and the use of compound piecewise linear measures was introduced by Wahlström [22]. We view all of these extensions under the umbrella of M&C, which has also found uses in the analysis of parameterized branching algorithms [11], whereas our piecewise analysis differs from M&C in a significant detail. We use a measure that not only depends on the (sub-)instance being solved by the current node of the search tree, but also on the original instance.

The remainder of this paper is structured as follows. Section 2 provides all necessary definitions and lemmas required by other sections. Section 3 describes the piecewise analysis method and its relation with M&C and Wahlström's compound measure. Following this, Section 4 presents our case study on $c$-Coloring and the application to 4-Coloring. The Section 5 presents the case study on #$c$-Coloring and the application to #3-Coloring. Finally, the conclusion summarizes our findings.

## 2     Preliminaries

Let $G = (V, E)$ be a simple undirected graph with vertex set $V$ ($|V| = n$) and edge set $E$ ($|E| = m$). Let $v \in V$ be a vertex in $G$. The *degree* of $v$ in $G$ is denoted by $d_G(v)$. The *maximum degree* of $G$ is denoted $\Delta(G)$. We denote the *open neighbourhood* of $v$ by $N(v) = \{u \in V : uv \in E\}$ and the *closed neighbourhood* by $N[v] = N(v) \cup \{v\}$. For a vertex subset $V' \subseteq V$, we use $G[V']$ to denote the subgraph of $G$ *induced* by $V'$. The set $V'$ is an *independent set of G* if $G[V']$ has no edge. The set $V'$ is a *vertex cover* of $G$ if $V \setminus V'$ is an independent set. The graph $G[V \setminus \{v\}]$ is also denoted $G - v$. $\mathcal{O}^*$-notation is similar to the usual $\mathcal{O}$-notation but allows hiding polynomial factors. Definitions of **Tree decomposition** and **Path decomposition**, along with related theorems, are provided in Appendix A.

To prove the correctness of the algorithm for $c$-Coloring in later sections, the following proposition is required.

▶ **Lemma 1** ([1]). *The* 3-*Coloring problem can be solved in time* $\mathcal{O}(1.3289^n)$ *for a graph on $n$ vertices.*

A new running time of $\mathcal{O}(1.3217^n)$ for 3-Coloring, claimed in [18], is not official accepted. If accepted, their result will further improve our 4-Coloring result by using it as a subroutine.[1]

---

[1] See more details in Subsection 4.3 and Theorem 18.

## 2.1  Measure & conquer

The M&C method was first introduced by Fomin et al.[10]. They highlighted a disproportionate focus on developing sophisticated algorithms over analysis methods within the research on exact exponential time algorithms.

Their measures allow for a fine-grained tracking of the progress a branching algorithm makes when solving an instance and allow us to amortize "slow" branching rules with "fast" ones. A measure that is too simple can easily fail to reflect progress made by branching rules, resulting in loose running time upper bounds. Fomin et al.[10] observed that a number of parameters, beyond the input size, are good candidates for measure. A measure $\mu$ for a problem $\Pi$ as a function $\mu : \mathcal{I} \to \mathbb{R}_{\geq 0}$ where $\mathcal{I}$ is the instance space (i.e., the set of inputs) of $\Pi$.

▶ **Lemma 2** ([12]). *Let $\mathcal{A}$ be a branching algorithm for a problem $\Pi$ whose search trees have polynomial depth, and let $\mu(\cdot)$ be a measure for the instances of $\Pi$, such that for any input instance $I$, $\mathcal{A}$ reduces $I$ to instances $I_1, ..., I_k$ in polynomial time, solves these recursively, and combines their solutions in polynomial time to solve $I$ and such that for any reduction done by Algorithm $\mathcal{A}$,*

$$(\forall I)\ 2^{\mu(I_1)} + ... + 2^{\mu(I_k)} \leq 2^{\mu(I)}\ . \tag{1}$$

*Then $\mathcal{A}$ solves any instance $I$ in $\mathcal{O}^*(2^{\mu(I)})$ time.*

Based on Lemma 2, we can obtain an upper bound on the running time of the algorithm $\mathcal{A}$, provided that all constraints required by the lemma are satisfied. In other words, the running time analysis becomes an optimization problem that minimizes an upper bound on the measure by finding suitable weights (see [14] and [12]) without violating any constraint imposed by the lemma.

## 3  Introduction to piecewise analysis

In this section, we introduce the main contribution of this paper: *piecewise analysis*, a new tool to analyze the worst-case running time of branching algorithms.

The intuition behind piecewise analysis is to group instances based on similarity and then analyze each group separately. To do this, we define a single similarity function, termed similarity ratio, that assigns a value to each instance. Then, instead of analyzing an algorithm $\mathcal{A}$ on the instance space $\mathcal{I}$ as a whole, we split $\mathcal{I}$ into *pieces* $P \subseteq \mathcal{I}$, each corresponding to a specific range of similarity ratio values. Then we analyze the worst-case running time of instances within each individual piece. Among all pieces, the one with the maximum value gives the upper bound of the running time of the algorithm $\mathcal{A}$ to $\mathcal{I}$.

This description naturally prompts the question: how do we construct a good similarity ratio? The short answer is to explore relations between the existing parameters that are relevant to the problem or the algorithm being analyzed.

The seemingly simple idea behind piecewise analysis holds potential. To showcase its potency, we apply it to analyze two 17-year-old algorithms in [6] that solve *c*-Coloring and #*c*-Coloring respectively. By applying our new piecewise analysis method, we improve the running time analysis of the algorithms for 4-Coloring and #3-Coloring presented in [6]. The original analysis in their paper gave running times of $\mathcal{O}(1.7272^n)$ and $\mathcal{O}(1.6262^n)$ respectively for these algorithms, and our analysis improves these running times to $\mathcal{O}(1.7207^n)$ and $\mathcal{O}(1.6225^n)$. Our running time $\mathcal{O}(1.7207^n)$ marks an improvement in the best-known running time for the 4-Coloring problem in 17 years.[2]

Before we delve into more details, the formal definition of piecewise analysis is provided below.

---

**Piecewise Analysis Framework**

| | |
|---|---|
| Input: | instance space $\mathcal{I}$ for an algorithm $\mathcal{A}$, |
| | parameters $r: \mathcal{I} \to \mathbb{R}_{\geq 0}$, $q: \mathcal{I} \to \mathbb{R}_{\geq 0}$, |
| | a finite cover $\mathcal{C} = \{P_0, P_1, ... P_p\}$ of the instance space where $P_0 = \{I \in \mathcal{I} : r(I) = 0\}$ and $p$ is a positive integer, |
| | lower and upper bounds $l, u : \mathcal{C} \setminus P_0 \to \mathbb{R}_{\geq 0}$ such that for all pieces $P \in (\mathcal{C} \setminus P_0)$ and any instance $I \in P$, we have $l(P) \leq \frac{q(I)}{r(I)} \leq u(P)$. |
| Requirement: | If $r(I) = 0$, then $I$ can be solved by $\mathcal{A}$ in polynomial time. |
| Oracle: | For each piece $P \in (\mathcal{C} \setminus P_0)$, the oracle uses $l(P), u(P)$ and parameter $r$ to generate an upper bound on the running time $\mathcal{O}^*(\gamma_P^r)$ for all of the instances in the piece, where $\gamma_P \in \mathbb{R}_{>0}$ is a constant. |
| Output: | Running time upper bound $max_{1 \leq i \leq p}\{\mathcal{O}^*(\gamma_{P_i}^r)\}$ for $\mathcal{A}$ over $\mathcal{I}$. |

---

### 3.1 Details of the piecewise analysis framework

The framework analyzes the running time for algorithm $\mathcal{A}$ on instance space $\mathcal{I}$ in terms of a parameter $r$. We assume that instances whose parameter $r$ is 0, denoted $P_0 = \{I \in \mathcal{I} : r(I) = 0\}$, are trivial and can be solved in polynomial time. For other instances in $\mathcal{I} \setminus P_0$, we divide them into pieces and select an auxiliary parameter $q$ to form the similarity ratio $\frac{q}{r}$. For each instance, this similarity ratio must be a real number that falls within an interval $[l_\mathcal{I}, u_\mathcal{I}]$, where $l_\mathcal{I}, u_\mathcal{I} \in \mathbb{R}_{\geq 0}$. We then split this interval into segments, each of which is of the form $[l, u]$ such that $l \leq u$. For a segment $[l, u]$, a *piece* is defined as the set of all instances whose similarity ratio falls within $[l, u]$. Since the union of all segments covers the whole interval $[l_\mathcal{I}, u_\mathcal{I}]$, the union of $\{P_1, ..., P_p\}$ and $P_0$ form the instance cover $\mathcal{C}$ of the instance space $\mathcal{I}$.

The oracle is an external source that takes all the inputs mentioned and obtains a running time upper bound in terms of $r$ for each piece $P \in (\mathcal{I} \setminus P_0)$. In practice, it can be any legitimate analysis that obtains a running time upper bound in terms of $r$.

The output of the framework is the maximum value among the running times of all pieces. Note that the piece $P_0$ does not affect this running time in the $\mathcal{O}^*$ notation.

### 3.2 Comparison with measure & conquer

When we apply M&C, we define a single measure $\mu$ for the entire instance space $\mathcal{I}$. By minimizing an upper bound for $\mu$, we conduct a single analysis by computing one set of weights. This means that these weights are uniform across all instances, and this uniformity marks a key difference from our piecewise analysis.

The main impetus for this work is to allow different weights for different instances. Thus, in piecewise analysis, by defining a similarity ratio, we allow a different analysis for instances with different ratios. This results in multiple measures and multiple analyses. Furthermore, by dividing an instance space $\mathcal{I}$ into pieces, we can incorporate additional information provided by the pieces to better upper bound the measure.

---

[2] The recently accepted ESA 2024 paper [23], not yet available at the time of writing, claims a better running time for 4-Coloring using a different method than that presented here.

M&C can be integrated into the piecewise analysis framework to analyze the running time of instances within each piece. In fact, if we apply piecewise analysis by setting the number of pieces to $p = 1$ and use the M&C analysis as the oracle, then this is equivalent to applying M&C on the entire instance space.

The M&C analysis within a piece $P$ can benefit from additional information about instances provided by the piece $P$. For example, when the similarity ratio falls within a restricted interval, this may reflect some structural property of all instances within the piece $P$ which we can exploit in our running time analysis.

### 3.3 Comparison with Walström's compound measure

Our work is not the first to explore different weights for different kinds of instances. Wahlström [21] uses a *compound measure* to allow measuring different kinds of instances in different ways.

His work using compound measures explores the idea that some intrinsic constraints are imposed by the properties of instances on various parameters. In [21], Wahlström gives the following example. If the average degree of $G$ is greater than $d$, then there is at least one vertex of degree of at least $d + 1$. He describes this kind of connection as implicit states. To model its effect in branching applicability, he uses a compound piecewise linear measure. However, similar to M&C, this approach uses a single, though compound, measure and still conducts only a single analysis, whereas in piecewise analysis we conduct multiple analyses.

A limitation of the compound measure is the need to ensure that the measure of an instance transitions smoothly between the components of the measure, since an instance may transition from being measured by one component to another. In piecewise analysis, an instance does not transition from one piece to another, since the measure is determined by the original input instance.

Similar to how M&C can be integrated in piecewise analysis, we can also use compound measure to analyze running time within each piece of an instance cover.

## 4 The $c$-Coloring problem – a case study

In this section, we explore how piecewise analysis can be used to analyze the running time of a $c$-Coloring algorithm in [6]. Our piecewise analysis of Fomin et al.'s algorithmic technique [6] improves the analysis of their 4-Coloring algorithm from $\mathcal{O}(1.7272^n)$ to $\mathcal{O}(1.7207^n)$, see Theorem 6.

In Subsection 4.1, we elucidate the framework in [6] and describe the algorithm and its pseudo code for $c$-Coloring. In Subsection 4.2, we combine piecewise analysis and measure & conquer to analyze the running time upper bounds of their algorithms. In Subsection 4.3, we showcase how applying piecewise analysis to 4-Coloring improves its running time compared to the one originally presented. In Subsection 4.4, we show how running time analysis improves with increased number of pieces of instance space, using 4-Coloring as an example.

> **$c$-Coloring**
>
> **Input:** A graph $G = (V, E)$
> **Question:** Can we color all the vertices of $G$ with at most $c$ colors so that no adjacent vertices share the same color?

▪ **Algorithm 1** `enumISPw`$(G, S, C)$

---

**Input**  : A graph $G$, an independent set $S$ of $G$, and a set of vertices $C$ such that $N(S) \subseteq C \subseteq V \setminus S$, integer $a \geq 3$, and $\alpha_i \in \mathbb{R}_{>0}$ for $i = 2, ..., a-1$
**Output**: $\bigvee$`enumIS`$(G, S', C')$ taken over all vertex covers $C' \supseteq C$ of $G$

**1  if** $(\Delta(G - (S \cup C)) \geq a)$
**2**     $\vee \big(\Delta(G - (S \cup C)) = a - 1 \text{ and } |C| > (\alpha_{a-1}n + \gamma_{a-1}|S|)\big)$
**3**     $\vee \big(\Delta(G - (S \cup C)) = a - 2 \text{ and } |C| > (\alpha_{a-2}n + \gamma_{a-2}|S|)\big)$
**4**     $\vee \cdots$
**5**     $\vee \big(\Delta(G - (S \cup C)) = 3 \text{ and } |C| > (\alpha_3 n + \gamma_3 |S|)\big)$
**6  then**
**7**   | choose a vertex $v \in V - (S \cup C)$ of maximum degree in $G - (S \cup C)$
**8**   | $T1 \leftarrow$ `enumISPw`$(G, S \cup \{v\}, C \cup N(v))$
**9**   | $T2 \leftarrow$ `enumISPw`$(G, S, C \cup \{v\})$
**10**  | **return** $T1 \vee T2$
**11 else if** $\Delta(G - (S \cup C)) \leq 2$ and $|C| > (\alpha_2 n + \gamma_2 |S|)$ **then**
**12**  | **return** `enumIS`$(G, S, C)$
**13 else**
**14**  | Stop this algorithm and run `Pw`$(G, S, C)$ instead

---

## 4.1 The algorithmic technique and the algorithm

The algorithmic technique in [6] solves *c*-Coloring by combining a pathwidth approach and an enumeration approach. Both approaches can be turned into independent exact algorithms for *c*-Coloring, but they shine on different kinds of graphs.

The pathwidth approach provides a running time upper bound $\mathcal{O}^*(c^w)$ where $w$ is the width of a path decomposition of the input graph $G$ and $c$ is the number of colors available. When $G$ has a small pathwidth, then this approach runs fast. The enumeration approach, on the other hand, enumerates (maximal) independent sets and then checks if the corresponding (minimal) vertex covers are $(c-1)$-colorable. Fomin et al.[6] demonstrated that the enumeration approach's worst-case instances have small pathwidth; moreover, it can detect these worst-case instances efficiently and hand over the computation to the pathwidth approach in these cases.

By exploiting this dichotomy and hybridizing these two approaches, graphs with nice graph-theoretic properties are differentiated from those with nice algorithmic properties. This distinction allows for the unique strength of each approach to be better utilized.

Algorithm 1 is adapted from [6] with minor modifications. Compared to [6], we pre-calculate an extra set of variables $\gamma_i$ for $2 \leq i \leq a - 1$ in line 2 to line 5 and line 11. These changes adjust the numeric thresholds for triggering the pathwidth approach without altering the application and combination of the two approaches. For clarity on whether the improvement comes from the modification or the application of piecewise analysis, see discussions at the end of Subsubsection 4.2.1 and in Subsection 4.4.

### 4.1.1 Details of the algorithm

The key step of Algorithm 1 is branching on a vertex of maximum degree $v \in G - (S \cup C)$ where $S$ is an independent set and $C$ is a vertex cover. The branching rule places $v$ into either $S$ or $C$, thus dividing the input $(G, S, C)$ into two branching pathways: $(G, S \cup \{v\}, C \cup N(v))$ and $(G, S, C \cup \{v\})$ respectively (line 7 to line 10).

7**Algorithm 2** enumIS$(G, S, C)$

**Input** : A graph $G$ with independent set $S$ and vertex set $C$, such that
$\Delta(G - (S \cup C)) \leq 2$ and $N(S) \subseteq C \subseteq V - S$
**Output:** $\bigvee \text{color}(G, C')$ taken over all vertex covers $C' \supseteq C$

1 **if** $\Delta(G - (S \cup C)) > 0$ **then**
2     choose a vertex $v \in V - (S \cup C)$ of maximum degree in $G - (S \cup C)$
3     $T1 \leftarrow \text{enumIS}(G, S \cup \{v\}, C \cup N(v))$
4     $T2 \leftarrow \text{enumIS}(G, S, C \cup \{v\})$
5     **return** $T1 \vee T2$
6 **else**
7     $S \leftarrow S \cup V(G - (S \cup C))$
8     run $\text{color}(G, C)$

This branching continues unless we hit one of the precalculated thresholds, upon which we then run a pathwidth algorithm Pw (line 14). Details on how to estimate the upper bound of pathwidth $w$, obtain the path decomposition, and calculate values $\alpha_i$ and $\gamma_i$ for $2 \leq i \leq a - 1$ can be found in Appendix D.

If Pw is executed, we obtain a path decomposition of $G$ efficiently and apply the pathwidth approach instead of continuing the enumeration approach. If the maximum degree of $G - (S \cup C)$ is 2 and Pw is not called, then enumIS (line 12) is called and continues to enumerate independent sets. It should be noted that the algorithm enumIS is flexible, and we may not need to explore the entire search space.

Algorithm 1, by starting enumerating independent sets $S$, can determine a pathwidth upper bound $w$ of the original input graph $G$. This enumeration has a twofold purpose: first, to apply the pathwidth approach if $w$ is small enough (line 1 to 5 and line 11); and second, to solve $c$-Coloring by using branching if $w$ is large.

### 4.1.2 Framework and details of subroutine enumIS

For enumIS, we employ a simplified version of the algorithm (see Algorithm 2) used in the paper[3]. It does the same two-way branching as enumISPw.

The input of enumIS is a graph $G$, an independent set $S$ and a vertex set $C$ such that $N(S) \subseteq C \subseteq V - S$ and $\Delta(G - (S \cup C)) \leq 2$. The algorithm arbitrarily chooses a vertex $v$ of maximum degree in $G - (S \cup C)$ and puts $v$ into either $S$ or $C$ (line 2 to line 5) until the maximum degree of $G - (S \cup C)$ drops to 0 (see line 1). When this happens, any vertex left in $G - (S \cup C)$ is added to $S$ (line 7) and then subroutine color is called (line 8). The algorithm color is an algorithm for $(c - 1)$-Coloring.

## 4.2 Running time analysis

In this subsection, we analyze the running time of the algorithm enumISPw (Algorithm 1) introduced in Subsection 4.1. It is important to note that analyzing running time for the pathwidth approach becomes straightforward once we obtain a pathwidth upper bound $w$ (see details in Appendix D). In contrast, the true challenge lies in analyzing the running time of the branching approach. In what follows, we will employ measure & conquer and piecewise analysis to analyze the running time of the branching approach in more detail.

---

[3] In [4], there are two branching rules on degree 2 vertex: rules on non-triangle cycles and rules on chains. This is slightly more complex than ours.

Before moving forward, we introduce a new problem, $d$-colorable Vertex Cover, to benefit the running time analysis of `enumISPw`.

---

**$d$-colorable Vertex Cover**

**Input:** A graph $G = (V, E)$ and non-negative integer $k \leq \frac{dn}{d+1}$
**Question:** Is there a vertex subset $X \subseteq V$ such that $G \setminus X$ is an empty graph, $|X| \leq k$, and $G[X]$ is $d$-colorable?

---

▶ **Proposition 3.** *A graph $G$ is $c$-colorable if and only if it has a $(c-1)$-colorable vertex cover $C$ with at most $\frac{(c-1)n}{c}$ vertices.*

**Proof.** see the proof in Appendix B ◀

▶ **Proposition 4.** *$c$-Coloring can be solved in $\mathcal{O}^*(\tau^n)$ time if and only if $(c-1)$-colorable Vertex Cover can be solved in $\mathcal{O}^*(\tau^n)$ time.*

**Proof.** see the proof in Appendix B ◀

As such, for the remainder of this section, our analysis of `enumISPw` focuses on $d$-colorable Vertex Cover.

### 4.2.1 Applying measure & conquer

In this subsection, we use measure & conquer to analyze Algorithm 1. Note that [6] did not use measure & conquer, so we need to first define a measure that can captures the progress made in each branching step. We define the following measure $\mu$ as

$$\mu = w_1(n - |S \cup C|) + \sum_{i=2}^{a-1} w_{ki} b_i + w_s b_s + c'k \tag{2}$$

where we set the parameters $b_i = \max(0, k - \alpha_i n - \gamma_i |S_i|)$ for all $2 \leq i \leq a - 1$ and, in turn, $|S_i|$ denotes the first value of $|S|$ in Algorithm 1 for which $\Delta(G - (S \cup C)) = i$. The intuition behind $b_i$ relates to the maximum number of vertices that can be added to $C$ when $\Delta(G - (S \cup C)) = i$. If the pathwidth approach is taken when $\Delta(G - (S \cup C)) = i$, then $b_i$ is set to 0 because branching rules (line 7 to 10) halt and `Pw` is invoked; if not, as referenced from line 1 to 5 and line 11 of Algorithm 1, we have $|C| > (\alpha_i n + \gamma_i |S|)$. This limits the subsequent additions to $C$ to a maximum of $b_i = k - \alpha_i n - \gamma_i |S|$ vertices. The parameter $b_s = n - k$ upper bounds the number of vertices added to the independent set $S$. We also include an extra term $c'k$, which corresponds only to the `color` subroutine. The constant $c'$ is based solely on `color` (see line 8 of Algorithm 2)[4]. For each parameter, we also define a weight variable: $w_1, w_s, w_{ki}$ for $2 \leq i \leq a - 1$.

The analysis now becomes a convex optimization problem (see Subsection 2.1) to compute the values for the weights to minimize an upper bound on the objective function $\mu$. Values

---

[4] In fact, `color` can be any algorithm that validly solves the $d$-Coloring problem with a running time of $\mathcal{O}^*(2^{c'n'})$ where $n'$ is the number of vertices of input graph.

assigned to the weights must satisfy all branching constraints, which can be found in Appendix E.

When the algorithm finishes enumerating all minimal vertex covers $C$, the $d$-Coloring algorithm `color` is called on each induced graph $G[C]$ of size at most $k$. The running time for the subproblem corresponding to each of our vertex covers of size $k$ is upper bounded by $\mathcal{O}^*(2^{c'k})$. Combining all of this, the overall running time for $d$-colorable Vertex Cover is $\mathcal{O}^*(2^{\mu})$ based on Lemma 2.

To upper bound $\mu$ in terms of $n$, we first upper bound it in terms of both $n$ and $k$ by the following function:

$$w_1 n + \sum_{i=2}^{a-1} w_{ki}(k - \alpha_i n) + w_s(n - k) + c'k \tag{3}$$

This is because $b_s = n - k$ by definition. Since the pathwidth approach is not triggered, the parameter $b_i \leq k - \alpha_i n$, because $b_i = k - \alpha_i n - \gamma_i |S_i|$ and $\gamma_i |S_i| \geq 0$ for $2 \leq i \leq a - 1$.

According to Proposition 3, if $G$ is $c$-colorable, then $0 \leq k \leq \frac{(c-1)n}{c}$. This allows us to further upper bound the measure $\mu$ by

$$w_1 n + \sum_{i=2}^{a-1} w_{ki}\left(\frac{(c-1)n}{c} - \alpha_i n\right) + w_s n + \frac{(c-1)c'n}{c} \tag{4}$$

The new upper bound is a running time in terms of only parameter $n$.

▶ **Example 5.** Using the above objective function (4) and constraints (8), (9) and (10) (see Appendix E), we can analyze 4-Coloring as follows. The subroutine `color` solves 3-Coloring in running time $\mathcal{O}(1.3289^n)$ (see Lemma 1), which means $c' = log_2(1.3289)$ because $\mathcal{O}^*(2^{c'n}) = \mathcal{O}(1.3289^n)$. By setting remainder of the variables to the values[5] in Table 5 (see Appendix F), we obtain a running time upper bound $\mathcal{O}(1.7275^n)$.

The running time presented for 4-Coloring in [6] is $\mathcal{O}(1.7272^n)$, which is very similar to the result we obtained in Example 5. Note that the result $\mathcal{O}(1.7275^n)$ comes only from applying M&C to `enumISPw`. This suggests that our previous modification on thresholds and the application of M&C do not significantly improve the analysis. As discussed in Subsection 3.2, applying M&C directly is equivalent to applying piecewise analysis seeing the instance space as one piece. The next section demonstrates that a larger number of pieces quickly improves the running time.

#### 4.2.2 Applying piecewise analysis

To apply piecewise analysis, it is essential to specify each input as required by the definition in Section 3.

The first input is the instance space of $d$-colorable Vertex Cover, denoted by $\mathcal{I}$. By definition of the problem, $\mathcal{I} = \{(G, k) | G \text{ is a graph, } k \in \mathbb{Z}_{\geq 0} \text{ and } k \leq \frac{dn}{d+1}\}$. The next two inputs $r$ and $q$ are chosen to be $n$ and $k$, respectively. Since we would like the final running time to be expressed in terms of $n$, we set $r := n$. If $n = 0$, then $G$ is a trivial yes-instance, which satisfies the requirement. Accordingly, we define $P_0 = \{(G, k) : G \text{ has no vertices, } k \in \mathbb{Z}_{\geq 0} \text{ and } k \leq \frac{dn}{d+1}\}$. The choice of $q$ is less obvious, but it is a natural choice since $k$ is at most linear in $n$, and the value $\frac{k}{n}$ falls nicely into $[0, \frac{d}{d+1}]$. With the $P_0$ and similarity ratio, we construct the instance cover in the following way. We *evenly* split the interval $[0, \frac{d}{d+1}]$ into $p$ segments $[l_i, u_i]$ for $i = 1, ..., p$, where $p \in \mathbb{Z}_{>0}$. For a segment $[l_i, u_i]$, we define a piece $P_i = \{(G, k) : G \text{ is a graph}, l(P_i) \leq \frac{k}{n} \leq u(P_i)\}$ where $l(P_i) = l_i = \frac{(i-1)(\frac{d}{d+1})}{p}$ and $u(P_i) = u_i = \frac{i(\frac{d}{d+1})}{p}$. As introduced in Section 3, since the union of all segments is the whole interval $[0, \frac{d}{d+1}]$, the union of pieces $\{P_1, ...P_p\}$ and $P_0$ form the instance cover $\mathcal{C}$ of the instance space $\mathcal{I}$. If we conduct a running time analysis on each piece, we cover the whole instance space $\mathcal{I}$. We ignore $P_0$ from now on as these instances are solved in constant time.

---

[5] The values are computed using AMPL with <u>this code</u>.

The intuition behind this construction is to group instances $(G, k)$ according to the closeness of their similarity ratio. With the upper and lower bounds of similarity ratio in each piece, we can further upper bound the measure defined in Subsubsection 4.2.1, see Subsubsection 4.2.3.

---

**Piecewise Analysis on the algorithm `enumISPw`**

Input: instance space $\mathcal{I} = \{(G, k) | G \text{ is a graph}, k \in \mathbb{Z}_{\geq 0} \text{ and } k \leq \frac{dn}{d+1}\}$,
parameters $r((G, k)) := n$, $q((G, k)) := k$,
a cover $\mathcal{C} := \{P_0, P_1, ..., P_p\}$ where $P_0 = \{(G, k) : G \text{ has no vertices}, k \in \mathbb{Z}_{\geq 0} \text{ and } k \leq \frac{dn}{d+1}\}$ and $P_i = \{(G, k) : G \text{ is a graph with } n > 0 \text{ and } \frac{(i-1)(\frac{d}{d+1})}{p} \leq \frac{k}{n} \leq \frac{i(\frac{d}{d+1})}{p}\}$ for $i = 1, ..., p$ for $p \in \mathbb{Z}_{>0}$,
lower bound $l(P_i) = \frac{(i-1)(\frac{d}{d+1})}{p}$ for $i = 1, ..., p$,
upper bound $u(P_i) = \frac{\frac{id}{d+1}}{p}$ for $i = 1, ..., p$,

Requirement: If $n((G, k)) = 0$, then the input instance is a yes-instance.

Oracle: For each piece $P \in \mathcal{C} \setminus P_0$, the oracle uses $l(P), u(P)$ and parameter $n$ to generate a running time upper bound $\mathcal{O}^*((\gamma_P)^n)$ for all of the instances in the piece, where $\gamma_P \in \mathbb{R}_{>0}$ is a constant.

Output: Running time upper bound $max_{1 \leq i \leq p}\{\mathcal{O}^*((\gamma_{P_i})^n)\}$ for `enumISPw` on $\mathcal{I}$.

---

### 4.2.3 Upper bound measure within a piece

To upper bound the measure $\mu$ (see Equation 2), it is sufficient to upper bound Equation 3, recalled here: $w_1 n + \sum_{i=2}^{a-1} w_{ki}(k - \alpha_i n) + w_s(n - k) + c'k$

Now we show how to upper bound Equation 3 within a piece. Given a piece $P \in (\mathcal{C} \setminus P_0)$, let $[l, u]$ be an interval containing all similarity ratio values of instances in $P$. Assuming $P \in (\mathcal{C} \setminus P_0)$ is the piece that $(G, k)$ belongs to, we have $\frac{k}{n} \in [l, u]$. Thus, we can establish that $k \leq un$ and $-k \leq -ln$. Now we can upper bound Equation 3 by

$$w_1 n + \sum_{i=2}^{a-1} w_{ki}(un - \alpha_i n) + w_s(n - ln) + c'un \quad (5)$$

Based on Lemma 2, the running time upper bound for instances in piece $P$ is $\mathcal{O}^*(2^\mu)$. Substituting $\mu$ by Equation 5 thus gives $P$ the below running time upper bound:

$$\mathcal{O}^*\left(2^{n\left(w_1 + \sum_{i=2}^{a-1} w_{ki}(u - \alpha_i) + w_s(1 - l) + c'u\right)}\right) \quad (6)$$

### 4.3 Application to 4-Coloring

▶ **Theorem 6.** *4-Coloring can be solved in $\mathcal{O}(1.7207^n)$ time.*

| Variable | Value | Variable | Value | Variable | Value | Variable | Value |
|---|---|---|---|---|---|---|---|
| $w_1$ | 0.35115 | $w_{k6}$ | 0 | $w_{k5}$ | 0 | $w_{k4}$ | 0.05121 |
| $w_{k3}$ | 0.07153 | $w_{k2}$ | 0.11110 | $w_s$ | 0.06390 | $\alpha_6$ | $-0.24466$ |
| $\alpha_5$ | $-0.07382$ | $\alpha_4$ | 0.08725 | $\alpha_3$ | 0.26980 | $\alpha_2$ | 0.39150 |
| $l$ | 0.74863 | $u$ | 0.75 | | | | |

**Table 1** 4-Coloring running time on 550 pieces

| #Pieces | 1 | 10 | 50 | 100 |
|---|---|---|---|---|
| Running time | $\mathcal{O}(1.7275^n)$ | $\mathcal{O}(1.7257^n)$ | $\mathcal{O}(1.7217^n)$ | $\mathcal{O}(1.7212^n)$ |

**Table 2** 4-Coloring running time on 10, 50, 100 pieces

**Proof.** By setting $p = 550$, $a = 7$, we obtain an instance cover of 550 pieces. Among all pieces, the one with the maximum running time is $P_{550} = \{(G, k) | G \text{ is a graph}, l \leq \frac{k}{n} \leq u\}$ where $l = 0.74863$ and $u = 0.75$. The subroutine `color` solves 3-Coloring in $\mathcal{O}(1.3289^n)$ time (see Lemma 1), which means $c' = log_2(1.3289)$ because $\mathcal{O}^*(2^{c'n}) = \mathcal{O}(1.3289^n)$.

Using upper bound (6) and constraints (8), (9) and (10) (see Appendix E), we can obtain the upper bound

$$\mathcal{O}^*\left(2^{n\left(w_1 + \sum_{i=2}^{7} w_{ki}(u-\alpha_i) + w_s(1-l) + c'u\right)}\right) = \mathcal{O}(1.7207^n)$$

using the values[6] in Table 1. ◀

### 4.4 Number of pieces

In applying piecewise analysis, we can decide how many pieces to divide the instance space $\mathcal{I}$ into. In the case study of 4-Coloring, we observe that the more pieces we divide the instance space $\mathcal{I}$ into, the tighter the running time we obtain. Table 2 provides four running time upper bounds for 4-Coloring when we set the number of pieces to 1, 10, 50, and 100, respectively[7]. The number of pieces is a hyper-parameter. All analyses are valid and eventually converge.

Compared to the running time $\mathcal{O}(1.7272^n)$ given in [6], applying M&C alone improves the upper bound by 0.0003, while piecewise analysis improves it by $1.7272 - 1.7257 = 0.0015$ with $p = 10$. Eventually, the value converges and we achieve a progress of $1.7272 - 1.7217 = 0.0055$ without any essential change to the algorithm.

## 5 The #$c$-Coloring problem – a case study

In contrast to the decision problem $c$-Coloring, the counting version #$c$-Coloring instead is to count the number of $c$-colorings of an input graph. Despite this difference, the [6] framework used in Section 4 also applies to #$c$-Coloring with some minor changes.

Our running time of the [6] algorithm to solve #$c$-Coloring is $\mathcal{O}(1.6225^n)$. This outperforms the running time $\mathcal{O}(1.6262^n)$ presented in the original paper. Note that our result for #3-Coloring is looser than the current record of $\mathcal{O}(1.5858^n)$ given in [24]. However, our findings highlight the improvement in the running time analysis due to the piecewise analysis method.

---

[6] The values are computed using AMPL, available <u>here</u>
[7] See Table 5, Table 6, Table 7, and Table 8 for variable values when the number of pieces $p$ is respectively 1, 10, 50 and 100.

We first describe the adapted algorithm and then, in a similar manner to Section 4, we conduct an analysis with M&C and piecewise analysis. Finally, our results for #3-Coloring are presented in Subsection 5.2.

## 5.1   The algorithm and analysis

The algorithms for #$c$-Coloring (see Algorithm 3 `enumISPwCounting` in Appendix C) are very similar to those for $c$-Coloring (Algorithm 1, `enumISPw`). The detailed descriptions, comparisons and pseudo code can be found in Appendix C. The analysis is also conducted in a similar manner to Section 4. We first define a new problem #$d$-colorable Vertex Cover (see Appendix G) and then prove that the Algorithm 3 solves the #$d$-colorable Vertex Cover in Proposition 7.

▶ **Proposition 7.** *#$c$-Coloring can be solved in $\mathcal{O}^*(\tau^n)$ time if and only if #$(c-1)$-colorable Vertex Cover can be solved in $\mathcal{O}^*(\tau^n)$ time.*

**Proof.** see the proof in Appendix B                                                                                                ◀

Next, we define a measure $\mu$ for `enumISPwCounting` and then upper bound its running time by applying piecewise analysis. Here, the measure $\mu$ is defined as

$$\mu = w_1(n - |S \cup C|) + \sum_{i=2}^{a-1} w_{ki}b_i + w_s b_s + c'k \tag{7}$$

where all variables and weights are defined the same way as in Section 4. For branching constraints, all (8), (9), and (10) (see Appendix E) still apply. The key difference is an extra branching constraint (see Equation 11 in Appendix E) for the branching rule on an isolated vertex $v$ in $G - (S \cup C)$ (see line 3 and line 4 of Algorithm 4).

The application of piecewise analysis to this extended set of branching rules mirrors the analysis in Subsubsection 4.2.2 and Subsubsection 4.2.3, and so is omitted for brevity.

## 5.2   Application to #3-Coloring

▶ **Theorem 8.** *Let $G$ be a graph on $n$ vertices, then the #3-Coloring can be solved in running time $\mathcal{O}(1.6225^n)$.*

**Proof.** see the proof in Appendix B                                                                                                ◀

## 6   Conclusion

The main contribution of this paper is a new method for the running time analysis of branching algorithms called piecewise analysis. It enables us to differentiate instances by defining a similarity ratio. This gives us the freedom to decide what instances should be analyzed in a similar way. And more importantly, which instances we should analyze differently. Using $c$-Coloring and #$c$-Coloring as two case studies, we showcased how piecewise analysis can be applied. In doing so, we also obtained faster running times for 4-Coloring and #3-Coloring without essential change of two 17-year old algorithms. An additional use of piecewise analysis is to identify the set of instances that determine the bottleneck of the running time analysis. Based on a defined similarity ratio, piecewise analysis identifies the piece with the worst running time upper bound. Thus, improving the analysis or the algorithm for the instances in this piece directly leads to improved running times.

## A  Definitions and theorems about tree decomposition

**Tree decomposition.** For a graph $G = (V, E)$, a *tree decomposition* is a pair $(T, X)$ where $T = (I, F)$ is a tree with $I$ as its node set and $F$ as its edge set and $X = \{X_i \subseteq V : i \in I\}$ is a set of *bags* associated with the nodes of $T$, with the following properties:
1. $\bigcup_{i \in I} X_i = V$,
2. for each edge $\{u, v\} \in E$, there exists a node $i \in I$ such that $\{u, v\} \subseteq X_i$, and
3. for each vertex $v \in V$, the sub-forest $T[\{i \in I : v \in X_i\}]$ is a tree.

The *width* of a tree decomposition is its largest bag size minus one. The *treewidth* of $G$ (introduced by [19]) is the smallest width among all tree decompositions of $G$, and is denoted $tw(G)$.

**Path decomposition.** A *path decomposition* is a tree decomposition $(T, X)$ that requires the tree $T$ to be a path. The *width* of a path decomposition is the maximum bag size minus one. The *pathwidth* of a graph $G$ is the smallest width among all path decompositions of $G$ and is denoted $pw(G)$. It is clear that $tw(G) \leq pw(G)$ for any graph $G$.

▶ **Proposition 9** ([2]). *If $G$ has degree at most 2, then $tw(G) \leq 2$.*

▶ **Proposition 10** ([7, 3]). *For any graph $G$,*

$$pw(G) \leq \frac{n_3}{6} + \frac{n_4}{3} + \frac{13n_5}{30} + \frac{23n_6}{45} + n_{\geq 7} + o(n)$$

*where $n_i$ denotes the number of vertices of degree $i \in \{3, 4, 5, 6\}$ in $G$. The parameter $n_{\geq 7}$ denotes the number of vertices of degree at least 7. Moreover, we can find a path decomposition of $G$ of the corresponding width in polynomial time.*

It should be noted that this result has been generalized to all degrees by [3], providing more precise pathwidth upper bounds for higher degrees.

▶ **Lemma 11** ([6]). *Given a graph $G = (V, E)$ with a tree decomposition of $G$ of width $w$, #c-Coloring can be solved in $\mathcal{O}^*(c^w)$ time.*

## B  Proofs

▶ **Proposition** (3). *A graph $G$ with vertex set $V$ is c-colorable if and only if it has at least one $(c-1)$-colorable (minimal) vertex cover $C$ with at most $\frac{(c-1)n}{c}$ vertices.*

**Proof.** To prove the forward direction, suppose $G$ is $c$-colorable. Let $\{P_1, ..., P_c\}$ be a $c$-Coloring of $G$ and $P_c$ be the largest color class. Thus $|P_c| \geq \frac{n}{c}$, otherwise $P_i < \frac{n}{c}$ for $i = 1, ..., c$ and $n = \sum_{i=1,...,c} |P_i| < n$, a contradiction. We can prove that $C = \bigcup_{i=1}^{c-1} P_i$ is a $(c-1)$-colorable vertex cover of $G$ of size at most $\frac{(c-1)n}{c}$. This implies that the graph $G$ and any nonnegative integer $k \leq \frac{(c-1)n}{c}$ is a yes-instance to $(c-1)$-colorable Vertex Cover.

The backward direction can be proved by contra-positive and then by contradiction. Suppose $G$ is not $c$-colorable but there exists a $(c-1)$-colorable vertex cover $C$ of $G$ and $|C| \leq \frac{(c-1)n}{c}$. This gives a contradiction directly because by coloring the vertices in $C$ with $c-1$ colors and vertices in $V \setminus C$ the $c_{th}$ color, graph $G$ has a $c$-Coloring. ◀

▶ **Proposition** (4). *c-Coloring can be solved in $\mathcal{O}^*(\tau^n)$ time if and only if $(c-1)$-colorable Vertex Cover can be solved in $\mathcal{O}^*(\tau^n)$ time.*

| Variable | Value | Variable | Value | Variable | Value | Variable | Value |
|---|---|---|---|---|---|---|---|
| $w_1$ | 0.27135 | $w_{k5}$ | 0.03089 | $w_{k4}$ | 0.04030 | $w_{k3}$ | 0.05501 |
| $w_{k2}$ | 0.75603 | $w_s$ | 0.58989 | $\alpha_5$ | 0.01268 | $\alpha_4$ | 0.16078 |
| $\alpha_3$ | 0.32863 | $\alpha_2$ | 0.44052 | $l$ | 0.66653 | $u$ | 0.66667 |
| $\alpha_6$ | $-0.14439$ | $w_{k6}$ | 0 | | | | |

**Table 3** #3-Coloring Running time on 5000 pieces

**Proof.** To prove both directions, it is sufficient to prove that a solution to *c*-Coloring gives a solution to $(c-1)$-colorable Vertex Cover in constant time and vice versa. This is proved by Proposition 3. ◂

▶ **Proposition** (7). *#c-Coloring can be solved in $\mathcal{O}^*(\tau^n)$ time if and only if $\#(c-1)$-colorable Vertex Cover can be solved in $\mathcal{O}^*(\tau^n)$ time.*

**Proof.** To prove both directions, it is sufficient to prove that the solution to #*c*-Coloring gives a solution to #*d*-colorable Vertex Cover in constant time and vice versa. Note that at least one color class of a *c*-Coloring has at least $\frac{n}{c}$ vertices. Call this color class $C_1$. The union of the remaining $c-1$ color classes always has at most $\frac{(c-1)n}{c}$ vertices.

The backward direction is straightforward, because once a $(c-1)$-colorable vertex cover $Y$ of size at most $\frac{(c-1)n}{c}$ is found, the remaining vertices in $V \setminus Y$ belong to $C_1$ and then we obtain a *c*-Coloring (see Proposition 3). This means the number of $(c-1)$-colorable vertex covers of size at most $\frac{(c-1)n}{c}$ is the same as the number of *c*-colorings.

For the forward direction, suppose $G$ has $p$ different *c*-colorings. For each *c*-coloring $(C_1 = Y_1, Y_2, \ldots, Y_c)$, where $|Y_c| \geq \frac{n}{c}$, taking the union $Y = \bigcup_{i=1}^{c-1} Y_i$ gives a unique $(c-1)$-colorable vertex cover $Y$ of $G$ of size at most $\frac{(c-1)n}{c}$. This means the number of $(c-1)$-colorable vertex covers of $G$ of size at most $\frac{(c-1)n}{c}$ is the number of *c*-colorings. ◂

▶ **Theorem** (8). *Let $G$ be a graph on n vertices, then the #3-Coloring can be solved in running time $\mathcal{O}(1.6225^n)$.*

**Proof.** By setting $p = 5000$ and $a = 7$, we obtain an instance cover containing 5000 pieces. Among all pieces, the one with the maximum running time is $P_{5000} = \{(G,k)|G$ is a graph, $\frac{4999}{7500} \leq \frac{k}{n} \leq \frac{2}{3}\}$. The subroutine `color` can count the number of 2-colorings in polynomial time, thus giving $c' = 0$.

Using upper bound (6) and constraints (8), (9), (10) (see Appendix E) and (11), we can obtain the upper bound

$$\mathcal{O}^*\left(2^{n\left(w_1 + \sum_{i=2}^{7} w_{ki}(u-\alpha_i) + w_s(1-l) + c'u\right)}\right) = \mathcal{O}(1.6225^n)$$

The values[8] assigned to variables are displayed in Table 3.

◂

---

[8] The values are computed using AMPL with this code.

## C  Algorithms for #c-Coloring

Similarly to enumISPw, the #$c$-Coloring algorithm enumISPwCounting also branches on a vertex of maximum degree $v \in G - (S \cup C)$ by placing it in $C$ or $S$. The description of the branching rule is the same as in enumISPw, however, there are three differences. Firstly, line 10 of enumISPw is an "or" operator while the line 10 of enumISPwCounting is a "plus" operator. Secondly, the subroutine Pw for #$c$-Coloring counts the number of $c$-colorings, instead of deciding whether a graph is $c$-colorable (see Lemma 11). Thirdly, the subroutine enumIS is replaced by enumISCounting (see Algorithm 4) in line 12. These subroutines enumIS and enumISCounting also have a few minor differences.

Firstly, while both enumISCounting and enumIS have the same branching rule, it is applied slightly differently: enumIS does not branch on vertices of degree 0 in $G - (S \cup C)$ while enumISCounting does[9]. For the decision problem $c$-Coloring, we directly add degree-0 vertices to $S$. But to solve #$c$-Coloring, we still need to consider these vertices when computing the number of $c$-colorings of $G$. Thus the algorithm branches on these vertices to see if they belong to an independent set of size at least $\frac{n}{c}$ or a vertex cover of size at most $\frac{(c-1)n}{c}$. Secondly, the color subroutine in enumISCounting solves the #$(c-1)$-Coloring problem while enumIS's color subroutine solves the decision problem.

**Algorithm 3** enumISPwCounting$(G, S, C)$

---
**Input**: A graph $G$, an independent set $S$ of $G$, and a set of vertices $C$ such that $N(S) \subseteq C \subseteq V - S$, integer $a \geq 3$, and $\alpha_i \in \mathbb{R}_{>0}$ for $i = 2, ..., a-1$
**Output**: The number of $c$-colorings of $G$

1 **if** $(\Delta(G - (S \cup C)) \geq a)$
2   $\vee\ (\Delta(G - (S \cup C)) = a - 1$ and $|C| > (\alpha_{a-1}n + \gamma_{a-1}|S|))$
3   $\vee\ (\Delta(G - (S \cup C)) = a - 2$ and $|C| > (\alpha_{a-2}n + \gamma_{a-2}|S|))$
4   $\vee \cdots$
5   $\vee\ (\Delta(G - (S \cup C)) = 3$ and $|C| > (\alpha_3 n + \gamma_3|S|))$
6 **then**
7 | choose a vertex $v \in V - (S \cup C)$ of maximum degree in $G - (S \cup C)$
8 | $T1 \leftarrow$ enumISPwCounting$(G, S \cup \{v\}, C \cup N(v))$
9 | $T2 \leftarrow$ enumISPwCounting$(G, S, C \cup \{v\})$
10 | **return** $T1 + T2$
11 **else if** $\Delta(G - (S \cup C)) \leq 2$ and $|C| > (\alpha_2 n + \gamma_2|S|)$ **then**
12 | **return** enumIS$(G, S, C)$
13 **else**
14 | Stop this algorithm and run Pw$(G, S, C)$ instead

---

[9] Line 7 in Algorithm 2 does not exist in Algorithm 4. Also, line 1 in Algorithm 4 branches on a degree-0 vertex in $G - (S \cup C)$.

**Algorithm 4** `enumISCounting`(G, S, C)

**Input** : A graph $G$ with independent set $S$ and vertex set $C$, such that
$\Delta(G - (S \cup C)) \leq 2$ and $N(S) \subseteq C \subseteq V - S$
**Output**: The number of $c$-colorings of $G$

1. **if** $\left|V\left(G - (S \cup C)\right)\right| > 0$ **then**
2.     choose a vertex $v \in V - (S \cup C)$ of maximum degree in $G - (S \cup C)$
3.     $T1 \leftarrow$ `enumISCounting`$(G, S \cup \{v\}, C \cup N(v))$
4.     $T2 \leftarrow$ `enumISCounting`$(G, S, C \cup \{v\})$
5.     **return** $T1 + T2$
6. **else**
7.     run `color`$(G, C)$

## D    How to calculate thresholds $\alpha$ and $\gamma$

This subsection explains how to calculate $\alpha$ and each $\alpha_i$ and $\gamma_i$ for $2 \leq i \leq a - 1$ where $a \in \mathbb{Z}$ and $a \geq 3$.

1. Let the input graph be $G = (V, E)$. After branching on all vertices of degree at least $i + 1$ for $2 \leq i \leq a - 1$, we obtain a partial solution $C$ and a potentially smaller instance $G' = G - (S \cup C)$ with maximum degree at most $i$.

2. Following the Proposition 10, we obtain a path decomposition $P' = (X'_1, X'_2, \ldots, X'_r)$ of $G'$ with a pathwidth upper bound $w_i = \beta_i n' + o(n')$ where $n' = |V'|$ and $\beta_i \in [0, 1)$. When $i = 2$, the constant $\beta_2 = 0$. Below example shows how to obtain $\beta_2$.

   ▶ **Example 12.** Let $G$ denote a graph and $\Delta(G) = 2$, hence the $pw(G)$ is a also constant $\leq 2$ because the tree decomposition of cycles, paths or isolated vertices is also a path decomposition. To put this back to the width formula $w_2 = \beta_2 n' + o(n')$, we get $\beta_2 = 0$ because $w_2$ is not determined by $n'$. The constant 2 plays no role in exponential running time and therefore omitted.

   When $3 \leq i \leq a - 1$, the constant $\beta_i$ can be found in the proposition. The below example shows how to obtain $\beta_i$ for $i = 4$.

   ▶ **Example 13.** Let $G$ denote a graph and $\Delta(G) = 4$. the Proposition 10 tells us:
   $$pw(G) \leq \frac{n_3}{6} + \frac{n_4}{3} + o(n)$$
   where $n_3$ and $n_4$ denotes the number of vertices of degree 3 and 4 in $G$, respectively. Since $n_3 + n_4 \leq n$, we have $pw(G) \leq \frac{n}{3} + o(n)$. This is how we obtain the constant $\beta_4 = \frac{1}{3}$.

3. Based on $P'$, we construct a path decomposition $P = (X_1, X_2, \ldots, X_r, X_{r+1}, \ldots, X_{r+|S|})$ of $G$ where $X_i = X'_i \cup C$ for every $1 \leq i \leq r$ and $X_{r+j} = \{v_j\} \cup C$ for every $v_j \in S$. $P$ has a pathwidth upper bound $w = |C| + w_i$ where $i = \Delta(G')$
   $$w = |C| + w_i = |C| + \beta_i n' + o(n') \leq |C| + \beta_i(n - |C| - |S|) + o(n)$$

   This leads to
   $$w \leq (1 - \beta_i)|C| + \beta_i n - \beta_i |S| + o(n)$$

4. If $w \leq (1 - \beta_i)|C| + \beta_i n - \beta_i |S| + o(n) \leq \alpha n$ for some $\alpha \in \mathbb{R}_{>0}$, then solving $c$-Coloring with path decomposition has running time upper bounded by $c^{\alpha n}$. This inequality gives us
   $$|C| \leq \frac{\alpha n - \beta_i n - o(n)}{1 - \beta_i} + \frac{\beta_i |S|}{1 - \beta_i} \leq \frac{(\alpha - \beta_i)n}{1 - \beta_i} + \frac{\beta_i |S|}{1 - \beta_i}$$

**Table 4**

| max degree $i$ | $\beta_i n' + o(n')$ | $\beta_i$ | $\alpha_i = \frac{\alpha - \beta_i}{1 - \beta_i}$ | $\gamma_i = \frac{\beta_i}{1 - \beta_i}$ |
| --- | --- | --- | --- | --- |
| 2 | 2 | 0 | $\alpha_2 = \alpha$ | $\gamma_2 = 0$ |
| 3 | $\frac{n'}{6} + o(n')$ | $\frac{1}{6}$ | $\alpha_3 = \frac{6\alpha - 1}{5}$ | $\gamma_3 = \frac{1}{5}$ |
| 4 | $\frac{n'}{3} + o(n')$ | $\frac{1}{3}$ | $\alpha_4 = \frac{3\alpha - 1}{2}$ | $\gamma_4 = \frac{1}{2}$ |
| 5 | $\frac{13n'}{30} + o(n')$ | $\frac{13}{30}$ | $\alpha_5 = \frac{30\alpha - 13}{17}$ | $\gamma_5 = \frac{13}{17}$ |
| 6 | $\frac{23n'}{45} + o(n')$ | $\frac{23}{45}$ | $\alpha_6 = \frac{45\alpha - 23}{22}$ | $\gamma_6 = \frac{23}{22}$ |

5. Let $\alpha_i = \frac{\alpha - \beta_i}{1 - \beta_i}$ and $\gamma_i = \frac{\beta_i}{1 - \beta_i}$. If maximum degree of $G'$ first drops to $i$ and $|C| \le \alpha_i n + \gamma_i |S|$, where $|S|$ is the number of vertices that are added to $S$ when $\Delta(G')$ first becomes $i$ (denote as $S_i$), then the pathwidth based algorithm solves the $c$-Coloring on $G$ in $c^{\alpha n}$.

6. Choice of $\alpha$: If $\alpha$ is too large, the *pathwidth* approach runs slowly since big $\alpha$ means high pathwidth. If $\alpha$ is chosen too small, then the pathwidth approach is underutilized, ended up with sparse graphs handled by branching approach. Thus, $\alpha$ acts as a decision point, determining the approach used. Our implementation lets the solver select the optimal $\alpha$ that satisfies all branching constraints.

▶ **Example 14.** Different choice of $\alpha$ value changes $\alpha_i$ and the final running time. Below shows some simple calculations on how a higher or lower $\alpha$ value influences $\alpha_4$ and running time.

- If $\alpha = 0.393$, then $\alpha_4 \frac{(3 \cdot 0.393 - 1)n}{2} = 0.0895n$. The running time is $\mathcal{O}(4^{0.393n}) = \mathcal{O}(1.7243^n)$. Since $\gamma_4 = \frac{1}{2}$, it should noted that when maximum degree of $G - (S \cup C)$ is 4, the pathwidth approach is triggered when $|C| \le 0.0895n + \frac{1}{2}|S|$.
- If $\alpha = 0.392$, then $\alpha_4 = \frac{3 \times 0.392n - n}{2} = 0.0881n$. The running time is $\mathcal{O}(4^{0.392n}) = \mathcal{O}(1.7219^n)$. In this setting, when maximum degree of $G - (S \cup C)$ is 4, the pathwidth approach is triggered when $|C| \le 0.0881n + \frac{1}{2}|S|$.

## E Constraints

The branching constraints below correspond to branching rules in line 8,9 of Algorithm 1 and line 3,4 of Algorithm 2, respectively [10].

1. If $\Delta(G - (S \cup C)) \ge a$, then select a vertex $v$ of highest degree. If $v$ is added to $S$, then $N(v)$ is added to $C$ and call `enumISPw(G, S ∪ {v}, C ∪ N(v))`. In this branch, $\mu$ decreases by at least $(a+1)w_1 + w_s + \sum_{j=2}^{a-1} \gamma_j w_{kj}$, because $n - |S \cup C|$ decreases by at least $a + 1$, $b_s$ decreases by at least 1. Each $b_j$ decreases by at least $\gamma_j w_{kj}$ where $j = 2, \ldots, a - 1$, because we need to update $S_i$ before $\Delta G'$ drops to $i$ for $2 \le i \le a - 1$. Otherwise, call `enumISPw(G, S, C ∪ {v})`. In this branch, $\mu$ decreases by at least $w_1$ because $n - |S \cup C|$ decreases by at least 1. The branching constraint is

$$2^{-\left((a+1)w_1 + \sum_{j=2}^{a-1} \gamma_j w_{kj} + w_s\right)} + 2^{-w_1} \le 1 \qquad (8)$$

2. If $\Delta(G - (S \cup C)) = i$ and $2 \le i \le a-1$, then select a vertex $v$ of highest degree $i$ ($i \ge 2$). If $v$ is added to $S$, then we add $|N(v)|$ vertices to $C$ and call `enumISPw(G, S ∪ {v}, C ∪ N(v))`.

---

[10] Implementation of objective function and constraints in AMPL are provided separately

| Variable | Value   | Variable | Value   | Variable   | Value     | Variable   | Value     |
|----------|---------|----------|---------|------------|-----------|------------|-----------|
| $w_1$    | 0.40391 | $w_{k4}$ | 0       | $w_s$      | 0         | $\alpha_4$ | 0.09150   |
| $w_{k3}$ | 0.07197 | $w_{k2}$ | 0.12023 | $\alpha_3$ | 0.27320   | $\alpha_2$ | 0.39433   |
| $w_{k5}$ | 0       | $w_{k6}$ | 0       | $\alpha_5$ | $-0.06883$ | $\alpha_6$ | $-0.23886$ |

**Table 5** 4-Coloring measure & conquer analysis

In this branch, $n - |S \cup C|$ decreases by at least $i + 1$, $b_s$ decreases by at least 1. For each $b_j = k - \alpha_j n - \gamma_j |S_j|$ where $j = i, ..., a-1$, we decrease $b_j$ by at least $i$ since $|N(v)| \geq i$. For each $b_j$ where $j \leq i-1$, we decrease $b_j$ by an extra $\gamma_j w_{kj}$ since we add a vertex to $S$ (and hence each $S_j$ increase by 1). In total, the measure $\mu$ decreases by $(i+1)w_1 + \sum_{j=i}^{a-1} i w_{kj} + \sum_{j=2}^{i-1} \gamma_j w_{kj} + w_s$. Otherwise, call `enumISPw`$(G, S, C \cup \{v\})$. This branch decreases $\mu$ by $w_1 + \sum_{j=i}^{a-1} w_{kj}$ because at least one vertex is added to $C$. The branching constraint is the following.

$$2^{-\left((i+1)w_1+\sum_{j=i}^{a-1} i w_{kj}+\sum_{j=2}^{i-1} \gamma_j w_{kj}+w_s\right)} + 2^{-\left(w_1+\sum_{j=i}^{a-1} w_{kj}\right)} \leq 1 \qquad (9)$$

3. If $\Delta(G - (S \cup C)) = 1$, then vertex $v$ is either added to $S$ or $C$. The former calls `enumIS`$(G, S \cup \{v\}, C \cup N(v))$ and the latter calls `enumIS`$(G, S \cup N(v), C \cup \{v\})$. Since $v$ and its neighbors are both degree 1 vertex, the measure decreases are the same for these two branches. The branching constraint is

$$2^{-(2w_1+\sum_{j=2}^{a-1} w_{kj}+w_s)} + 2^{-(2w_1+\sum_{j=2}^{a-1} w_{kj}+w_s)} \leq 1 \qquad (10)$$

4. If $\Delta(G - (S \cup C)) = 0$, then vertex $v \in G - (S \cup C)$ is either added to $S$ or $C$. The former calls `enumISCounting`$(G, S \cup \{v\}, C)$ and the latter calls `enumISCounting`$(G, S, C \cup \{v\})$. The branching constraint is

$$2^{-(w_1+w_s)} + 2^{-(w_1+\sum_{j=2}^{a-1} w_{kj})} \leq 1 \qquad (11)$$

## F    Running time analysis examples for 4-Coloring

▶ **Example** (5). *Using the objective function* (4) *and constraints* (8), (9) *and* (10) *in Appendix E, we can analyze* 4-*Coloring as follows. The subroutine* `color` *solves* 3-*Coloring in running time* $\mathcal{O}(1.3289^n)$ *(see Lemma 1), which means* $c' = \log_2(1.3289)$ *because* $\mathcal{O}^*(2^{c'n}) = \mathcal{O}(1.3289^n)$. *By setting remainder of the variables to the values*[11] *in Table 5, we obtain a running time upper bound* $\mathcal{O}(1.7275^n)$.

▶ **Example 15.** By setting $p = 10$, $a = 7$, we obtain an instance cover of 10 pieces. Among all pieces, the one with the maximum running time is $P_{10} = \{(G, k) | G \text{ is a graph}, l \leq \frac{k}{n} \leq u\}$ where $l = 0.675$ and $u = 0.75$. The subroutine `color` solves 3-Coloring in running time $\mathcal{O}(1.3289^n)$ (see Lemma 1), which means $c' = \log_2(1.3289)$ because $\mathcal{O}^*(2^{c'n}) = \mathcal{O}(1.3289^n)$.

Using the function (6), constraints (8),(9) and (10) , we obtain the running time $\mathcal{O}(1.7270^n)$ with the values in Table 6.

▶ **Example 16.** By setting $p = 50$, $a = 7$, we obtain an instance cover of 50 pieces. Among all pieces, the one with the maximum running time is $P_{50} = \{(G, k) | G \text{ is a graph}, l \leq \frac{k}{n} \leq u\}$ where $l = 0.735$ and $u = 0.75$. The subroutine color solves 3-Coloring in running time $\mathcal{O}(1.3289^n)$ (see Lemma 1), which means $c' = log_2(1.3289)$ because $\mathcal{O}^*(2^{c'n}) = \mathcal{O}(1.3289^n)$.

---

[11] The values are computed using AMPL with this code.

| Variable | Value | Variable | Value | Variable | Value | Variable | Value |
|---|---|---|---|---|---|---|---|
| $w_1$ | 0.35115 | $w_{k4}$ | 0.05121 | $\alpha_6$ | $-0.24043$ | $\alpha_3$ | 0.27228 |
| $w_{k6}$ | 0 | $w_{k3}$ | 0.07153 | $\alpha_5$ | $-0.07018$ | $\alpha_2$ | 0.39357 |
| $w_{k5}$ | 0 | $w_{k2}$ | 0.11110 | $\alpha_4$ | 0.09035 | $w_s$ | 0.06390 |
| $l$ | 0.675 | $u$ | 0.75 | | | | |

**Table 6** 4-Coloring running time on 10 pieces

Using the function (6), constraints (8),(9) and (10), we obtain the running time $\mathcal{O}(1.7224^n)$ with the values in Table 7.

| Variable | Value | Variable | Value | Variable | Value | Variable | Value |
|---|---|---|---|---|---|---|---|
| $w_1$ | 0.35115 | $w_{k4}$ | 0.05121 | $\alpha_6$ | $-0.24388$ | $\alpha_3$ | 0.27026 |
| $w_{k6}$ | 0 | $w_{k3}$ | 0.07153 | $\alpha_5$ | $-0.07315$ | $\alpha_2$ | 0.39188 |
| $w_{k5}$ | 0 | $w_{k2}$ | 0.11110 | $\alpha_4$ | 0.08782 | $w_s$ | 0.06390 |
| $l$ | 0.735 | $u$ | 0.75 | | | | |

**Table 7** 4-Coloring running time on 50 pieces

▶ **Example 17.** By setting $p = 100$, $a = 7$, we obtain an instance cover of 100 pieces. Among all pieces, the one with the maximum running time is $P_{100} = \{(G,k)|G \text{ is a graph}, l \leq \frac{k}{n} \leq u\}$ where $l = 0.7425$ and $u = 0.75$. The subroutine color solves 3-Coloring in running time $\mathcal{O}(1.3289^n)$ (see Lemma 1), which means $c' = log_2(1.3289)$ because $\mathcal{O}^*(2^{c'n}) = \mathcal{O}(1.3289^n)$.

Using the function (6), constraints (8),(9) and (10), we obtain the running time $\mathcal{O}(1.7219^n)$ with the values in Table 8.

| Variable | Value | Variable | Value | Variable | Value | Variable | Value |
|---|---|---|---|---|---|---|---|
| $w_1$ | 0.35115 | $w_{k4}$ | 0.05121 | $\alpha_6$ | $-0.24431$ | $\alpha_3$ | 0.27000 |
| $w_{k6}$ | 0 | $w_{k3}$ | 0.07153 | $\alpha_5$ | $-0.07352$ | $\alpha_2$ | 0.39167 |
| $w_{k5}$ | 0 | $w_{k2}$ | 0.11110 | $\alpha_4$ | 0.08750 | $w_s$ | 0.06390 |
| $l$ | 0.7425 | $u$ | 0.75 | | | | |

**Table 8** 4-Coloring running time on 100 pieces

## G  Problems

> **#$d$-colorable Vertex Cover**
>
> **Input:** A graph $G = (V, E)$ and non-negative integer $k \leq \frac{dn}{d+1}$
> **Question:** How many vertex subsets $X \subseteq V$ are there, such that $G \setminus X$ is an empty graph, $|X| \leq k$, and $G[X]$ is $d$-colorable?

## H   Others

▶ **Theorem 18.** 4-*Coloring can be solved in running time* $\mathcal{O}(1.7146^n)$ *if the* 3-*Coloring subroutine algorithm we applied has running time* $\mathcal{O}(1.3217^n)$ *[18].*[12]

---

[12] The work is unofficially published.